\documentstyle[12pt]{article}
\def\CH{{\cal H}}
\def\CBP{\bar{\cal P}}
\def\BC{\bar{C}}
\def\CP{{\cal P}}

\begin{document}

\title{BRST-anti-BRST covariant theory \\
for the second class constrained systems.\\
A general method and examples.}
\author{I.Yu. Karataeva
and S. L. Lyakhovich}
\date{{\it Department of Physics, Tomsk State
University,}\\
{\it Lenin Ave. 36, Tomsk 634050, Russia }}
\maketitle

\begin{abstract}

The BRST-anti-BRST covariant extension is suggested for the split
involution quantization scheme for the second class constrained
theories. The constraint algebra generating equations
involve on equal footing a pair of BRST charges for second class
constraints and a pair of the respective anti-BRST charges. Formalism
displays explicit $ Sp(2) \times Sp(2)$ symmetry property.
Surprisingly, the the BRST-anti-BRST algebra must involve a central
element, related to the nonvanishing part of the constraint
commutator and having no direct analogue in a first class theory.
The unitarizing Hamiltonian is fixed by the requirement of
the explicit BRST-anti-BRST symmetry with a much more restricted
ambiguity if compare to a first class theory or split involution
second class case in the nonsymmetric formulation.

The general method construction is supplemented by the explicit
derivation of the extended BRST symmetry generators for
several examples of the second class theories, including self--dual
nonabelian model and massive Yang Mills theory.

\end{abstract}

\section{Introduction}

The BFV--BRST theory originally appeared as a tool for
quantization of the gauge fields or, from the standpoint of a general
Hamiltonian formalism, constrained dynamical systems. Hamiltonian
BFV--BRST formalism first developed  by Batalin, Fradkin and
Vilkovisky \cite{BFV} (for review see \cite{BatFrad,HenTeit}) is
well established today to solve, in principle, quantization problem
in a general first class constrained system. The
anti-BRST symmetry \cite{old} attracted considerable interest due to
an elegant structure revealed in quantum theory possessing both BRST
and anti--BRST symmetry on equal footing
\cite{BLT1,BLT2,Hull,HenGreg}.  In the $Sp(2)$ covariant description,
BRST and anti-BRST charges form a doublet $\Omega^a$, $a=1,2$. In
nonsymmetric approach, one of them, say $\Omega^1$, could be
identified to a charge, and another one to an
anti--charge.  An $Sp(2)$ symmetry of the
formalism can be retained at all the stages of the quantization
procedure although it restricts the choice of gauge fixing term.

In contrast to a first class constraint theory, a second class
one does not have today a universal tool for quantization.
Three basic trends could be distinguished among the approaches to
quantization of second class constrained systems.

The first trend is
to reduce (directly or indirectly) the phase space eliminating the
constraints. This approach is often related to an attempt to find a
realization for a Dirac bracket in quantum theory. In a theory with
nonlinear constraints, this bracket may depend upon the fields and,
generally speaking, the bracket appears to be nonlocal in a local
theory.  So it is hardly possible to expect too much from this idea
from the standpoint of the operator quantization in the theories
with nontrivial second class constraints.  In the path
integral
formalism, the idea of reduction results in a singular measure
\cite{Frad,Senjan}.

The second trend, quite opposite to the
first one, implies to extend the initial phase space by auxiliary
variables to convert the original second class constraints into
effective first class ones in the extended manifold. This idea was
applied probably first by Stuckelberg \cite{Stuck} to gauge the
massive abelian vector
field theory introducing an auxiliary scalar
field.  From the viewpoint of the general Hamiltonian constrained
dynamics, the conversion idea was suggested in Ref \cite{FSh,BF} and
the respective general theory was developed in  Ref\cite{BFF,BT}.
Today we observe a fairly large number of attempts to exploit this
method (and the numerous reformulations) in various
applications including superparticles and superstrings, sigma models,
massive Yang-Mills fields, etc. The advantage of this approach is
that it allows to apply well established machinery of the Hamiltonian
BFV-BRST method as soon as one has converted the theory in the first
class. The weak point in the conversion idea is related to the
possible conflict between the space-time covariance and locality and
embedding of the original second class surface in the extended
manifold. What is more, in general the effective first class theory
is equivalent to the original second class one only locally in the
phase manifold, so some of the topologically determined properties of
the second class theory could be lost when the conversion is
performed.

The third trend is an attempt to quantize second class
theory as it is, i.e. without recourse to reduction or extension of
the original phase space. With this regard we refer to the papers
\cite{split1,split2,mar1,mar2}. In particular, in the split
involution approach \cite{split1,split2}, the second class theory is
quantized with the original phase space variables (no extra variables
introduced) subject to {\it canonical} commutation relations, whereas
the constraints are accounted by weak conditions selecting physical
states. These conditions are related to the extended BRST symmetry
associated in this approach to the second class constraints.  This
BRST symmetry is generated for the pure second class constrained
theory \cite{split1} by a
pair of independent BRST charges constructed from the  original
constraints\footnote{Generalization of the method to the
case of both first and second class is given in \cite{split2},
the existence theorem for this extended BRST algebra is proven in
\cite{split3}}.
The only restriction is that the specific constraint
basis should be chosen for this construction to obey the split
involution conjecture \cite{split1} (which is briefly recalled in the
next section). These two charges form $Sp(2)$ doublet, and the
 split involution formalism displays certain resemblance to the
of the BRST-anti-BRST $Sp(2)$ covariant formulation of the gauge
theory. However one should remember of the important distinctions as
well: 1) in the split involution case both the charges are  BRST
ones, i.e., they have the ghost number 1, whereas in the
BRST-anti-BRST theory charges have the opposite ghost numbers;
2)  In split involution case each of the charges includes its own
set of the constraints, being independent from the set of another
charge involved in the doublet, in BRST-anti-BRST theory both charges
include the same constraints; 3) the split involution BRST symmetry
is not related to any gauge invariance of the original constrained
theory, as it is of the second class, whereas BRST and anti-BRST
symmetry in the first class theory could be thought about as a
residual global invariance emerged from the original gauge
transformations. Thus split involution approach allows to construct
directly  an extended BRST symmetric description for the second class
theory without reduction or extension of the original theory.

\vspace{5mm}

In this paper, we suggest the BRST-anti-BRST symmetric
extension for the split involution method for a pure second class
constrained theory. This extension shows some essentially new
features both from the viewpoint of the algebra and it's dynamical
content as compared to the respective construction known in a first
class theory. In particular, we observe that the BRST-anti-BRST
algebra must have a central extension in the second class theory.
The central element originates from the revertible element in the second
class constraint commutators. This extended formulation allows to
explicitly construct BRST-anti-BRST invariant quantum theory for any
second class system if the constraint basis was found subject to the
split involution conjecture. Stress once again that, in our approach,
{\it the BRST embedding of the second class system is performed
without transformation of the second class constraints into effective
first class ones and without any auxiliary variables "gauging" the
original theory.}  To exemplify the general method, we consider two
systems where the split involution constraint basis can be explicitly
found.  Both these second class theories  are embedded in the
BRST-anti-BRST symmetric theory as they are, without any converting
procedure.

One of the examples is a massive Yang-Mills field theory in $d=4$
which is a traditional test model for polishing out the general
schemes for the quantization second class systems.  Another
considered example  is the self--dual nonabelian
 model \cite{TPN} which is
extensively discussed in the last few years in relation to the BRST
quantization of the second class systems. The usual treatment of the
model (see \cite{TPNCo} and references therein) employs the idea to
embed the original second class theory in a more wide phase space of
the first class system possessing an artificial gauge invariance. It
is the gauge symmetry which underlies the BRST invariance in the
converting approach.
This effective BRST invariance of the converted first class theory
breaks the explicit self-duality of the model.  The advantage of our
approach in this case, besides simplicity, is that it does not cause
any contradiction between explicit self--duality and BRST symmetry.

\vspace{0.5cm}

The paper is organized as follows. The next section briefly recalls
the basic features of the split involution construction without
explicit anti-BRST symmetry. In Section 3 we introduce the
ghost-anti-ghost covariant notations and define the BRST-anti-BRST
invariant extension for the generating equations for the second class
constraint algebra.
Then we find an explicit solution to the generating
equations for the case when the split involution relations are
restricted to define Lie algebras.
Section 4 contains discussion of the unitarizing
Hamiltonian defined in  an explicitly BRST-anti-BRST covariant form.
In the end of this Section we show how to reproduce the non-extended
formulation \cite{split1} within the BRST-anti-BRST symmetric
approach.
In Section 5 we consider apply the general scheme to specific models
to exemplify the method.
 In conclusion we discuss some of the possible applications of the
 paper results and some of their peculiarities.

\section{Split involution construction.}

Consider two sets of linearly independent second class constraints
$ T^{a}_{\alpha}(q,p)$ , \,\,
$a=1,2\, ; \  \alpha = 1,...,m$.\footnote{ In this paper we consider
only even Grassmann
constraints to avoid sign factors, although no essential problem
appears in an odd case as well. } The Poisson bracket of the
constraints has symmetric and antysymmetric parts in $a \, b$ :
\begin{equation}
\{T^a_\alpha , T^b_\beta \}=\frac{1}{2}\{T^{(a}_\alpha ,T^{b)}_\beta
\}+ \frac{1}{2}\{T^{[a}_\alpha ,T^{b]}_\beta \} \, ,
\label{constrcom} \end{equation}
$$
A^{\{a \, b\}}= A^{(a\,b)}= A^{ab}+A^{ba}\, ,  \quad A^{[ab]}=
A^{ab}-A^{ba} \, ,
$$
\begin{equation}
\{T^{[a}_\alpha ,T^{b]}_\beta
\}\equiv \epsilon^{ab}\Delta_{\alpha\beta}\, .  \label{antisym}
\end{equation}
where $\varepsilon_{ab}$ is an $Sp(2)$--invariant
constant tensor \,
$
\epsilon^{12}=1,\ \ \  \epsilon_{12}=-1,\ \ \  \epsilon^{ab}\epsilon_{bc}=\delta^a_c.
$

The split involution conjecture \cite{split1} implies that it is the
antisymmetric part (\ref{antisym}) which forms
the invertible element in the constraint matrix (\ref{constrcom})
\begin{equation}
\exists \Delta^{-1}\, , \qquad
\Delta_{\alpha\beta}\equiv\Delta_{\beta\alpha}\equiv
\epsilon_{ba}\{T^a_\alpha , T^b_\beta \} \,  ,
\label{Delta}
\end{equation}
The symmetric part of the constraint commutator (\ref{constrcom}) and
the Hamiltonian should obey the split involution relations
\begin{equation}
\{ \, T^{(a}_\alpha \,  , \, T^{b)}_\beta \, \}=
U^{(a}_{\alpha\beta}{}^\gamma(q,p) \,\, T^{b)}_\gamma\, ,
\label{split}
\end{equation}
$$ U^{a}_{\alpha\beta}{}^\gamma = - \,
\, U^{a}_{\beta\alpha}{}^\gamma\, ,
$$
\begin{equation}
\{ \, H  \, , \, T^{b}_\beta \}=
V_\beta{}^\gamma T_\gamma^b \, , \quad V_\beta{}^\gamma=
V_\beta{}^\gamma \, (q,p)
\label{splitH}
\end{equation}
The split involution relations (\ref{split},\ref{splitH}) do not
actually restrict the second class constraint surface, although they
require a special basis for the constraints and a special choice of
the Hamiltonian outside the constraint surface. In the simplest case,
when $U^{a}_{\alpha\beta}{}^\gamma={\rm const}$, the relations
(\ref{split}) define a {\it pair of Lie algebras} mutually related
by a certain compatibility condition involving
both sets of the structure constants $U^1,\,U^2$ \cite{split1}:
\begin{equation}
U^{(a}_{\alpha\beta}{}^\gamma U^{b)}_{\gamma\nu}{}^\rho+
U^{(a}_{\beta\nu}{}^\gamma U^{b)}_{\gamma\alpha}{}^\rho +
U^{(a}_{\nu\alpha}{}^\gamma U^{b)}_{\gamma\beta}{}^\rho =0
\, ,
\label{Jac}
\end{equation}
These conditions include, besides the Jacobi identity for $U^1$ and,
independently, $U^2$, some additional restrictions (appeared when
$a\neq b$ in (\ref{Jac})) to the respective Lie algebras.  For the
case $V_\beta{}^\gamma = const$, it should obey the compatibility
condition following immediately from (\ref{split}), (\ref{splitH}):
$$
U^a_{\alpha\beta}{}^\gamma V_\gamma {}^\nu - V_\alpha{}^\gamma
U^a_{\gamma\beta}{}^{\nu} + V_{\beta}{}^{\gamma} U^a_{\gamma\alpha}{}^{\nu}=0.
$$

To build the BRST invariant theory, the split involution
method implies to introduce Lagrange multipliers
$\lambda^a_\alpha \, , \, \, a=1,2 \, , \, \, \alpha=1,\ldots , m$
\begin{equation}
\{ \lambda^{a}_\alpha , \,  \lambda^{b}_{\beta} \}
= d_{\alpha \beta} \, \epsilon^{ab} \, , \quad
d_{\alpha\beta}\equiv d_{\beta\alpha}={\rm const} \,  , \quad \exists
d^{\alpha\beta}\, :  \, \, d^{\alpha\beta}d_{\beta\gamma}=
\delta^\alpha_\gamma \, ,
\label{lambda}
\end{equation}
and an odd ghost variable set for each the pair $T^a_\alpha, \, a=1,2\,$:
\begin{equation}
\{ \, C^\alpha \, , \, \CBP_\beta \, \} =  \delta^\alpha_\beta  \, ,
\quad
\{ \, \CP^\alpha \, , \, \BC_\beta \, \} =  \delta^\alpha_\beta
\, , \quad  gh({C})=gh(\CP)=1 \, , \, gh({\BC})=gh({\CBP})=-1 \, .
\end{equation}
Next is to construct a $Sp(2)$ doublet of the BRST
generators $Q^a$ and BRST invariant Hamiltonian $\CH$, being
defined by the $Sp(2)$ covariant generating equations
\begin{equation}
\{\, Q^a \, , \,  Q^b\,  \} = 0\, , \quad \{ \, Q^a \, ,\CH \, \} = 0,
\label{QQQH}
\end{equation}
Then the complete Unitarizing Hamiltonian
of the theory is built in the $Sp(2)$--symmetric form
\begin{equation}
H_{complete}=\CH+ \frac{1}{2}\varepsilon_{ab} \, \{ \, Q^b \, ,
\, \{ \,  Q^a \, ,\, B \, \}\} \, ,
\label{QQB}
\end{equation}
where $B$ is known as a "gauge--fixing" Boson, $gh(B)=-2$.
The simplest possible choice for $B$ is as follows
\begin{equation}
B=\bar{{\cal P}}_\alpha \bar{C}^\alpha
\label{B}
\end{equation}
Being the physical quantities
defined in an invariant way, they do not depend on a particular
choice of the $B$. This independence is quite a nontrivial feature of
the split involution scheme \cite{split1,split2}, because we have pure
second--class constraints which do not generate an actual gauge
symmetry \footnote{ The relations (\ref{QQQH}, \, \ref{QQB}) coincide
at the first glance to the respective relations of  the formalism
developed in Ref \cite{BLT1,BLT2} to quantize gauge--invariant
theories formulated in a ghost--anti--ghost symmetric fashion.
However, the ghost numbers of the BRST generators ($Q^1$, $Q^2$) are
($+1$, $+1$) in the split involution scheme, while in the
ghost--anti--ghost symmetric first class theory these numbers are
($+1$, $-1$).  The "gauge" boson $B$ (\ref{QQB}) has ghost number
$-2$, in distinction to zero number of the respective quantity in the
BRST--anti--BRST gauge theory. }.

The solution for the Fermions $Q^a , \, \, a=1,2$ and Boson $\CH$ is
sought in the form of a series expansion in ghost powers
\begin{equation}
Q^a=C^\mu \, T^a_\mu \, + \, \CP^\alpha \,  \lambda^a_\alpha \, + \,
\ldots \, , \quad
\CH=H+\ldots \, .
\label{boundQH}
\end{equation}
These relations can be thought about as a boundary
condition to the generating equations (\ref{QQQH})
Mention that the Poisson bracket between $Q^a$ does not include
an antisymmetric combination $a$ and $b$, so the
commutativity of the charges (\ref{QQQH}) has no contradiction to the
noncommutativity of the corresponding constraints (\ref{Delta}).

The relations (\ref{boundQH}) show again the distinction between
split involution scheme and $Sp(2)$ covariant BRST-anti-BRST
formulation of a first class theory: in the latter case, the ghosts
and anti ghosts form an $Sp(2)$ doublet entering the charges, while
in the split involution the constraints and Lagrange multipliers
introduce the $Sp(2)$ transformation space.

We would like to note again that
the extended BRST symmetry in the split involution theory
(\ref{QQQH},  \ref{QQB}, \ref{boundQH}) has no relation to any
gauge transformations as all the constraints $T^a_\alpha$, being
involved into $Q^a$ (\ref{boundQH}),  are of the second class indeed
(\ref{Delta}). The commutativity of the charges and Hamiltonian
(\ref{QQQH}) is possible in spite of noncommutativity of the
constraints themselves (\ref{Delta}).
The BRST symmetry emerges here not from the gauge identities of the
original theory, it follows from the remarkable polarization of the
second class constraint surface revealed in the split involution
relations (\ref{split}, \, \ref{splitH}).

\section{BRST--anti--BRST symmetric description \newline
for the second class theory}

In this section, we  introduce  an anti--BRST charge for each of the
charges $Q^a$ and formulate the BRST--anti--BRST symmetric extension
for the generating equations (\ref{QQQH}, \ref{boundQH}) of the split
involution method.

At first we need the covariant notation for the ghost anti-ghost
variables themselves.
Introduce a doublet of the odd Grassman ghost canonical pairs \,
$C^{\dot{a}\, \alpha}, \, \CP_{\dot{b}\,\alpha}$,  $\dot{a}=1,2$
for each the constraint pair $T^a_\alpha$.
The indices of the $Sp(2)$ group, which specify
 ghost or anti--ghost canonical pair in this doublet, are denoted by
dots in distinction from the indices related to the second class
constraints $T^a_\alpha$ and the corresponding Lagrange multipliers
(\ref{lambda})

\begin{equation} \big\{ \, C^{ \dot{a}\, \alpha}\,
, \, \CP_{\dot{b}\, \beta} \, \big\} = \delta^{\dot{a}} {}_{\dot{b}}\
\ \delta^{\alpha}{}_\beta \, ,  \label{CaPb}
\end{equation}
$$
gh(C^{\dot{a}\, \alpha}) = - gh(\CP_{\dot{a}\,\alpha})= (-
1)^{\dot{a}+1}.
$$

We find sometimes convenient to use both lower and
upper indices $\dot{a}=1,2$ and $\alpha =1,\ldots,m$:
$$
\CP^{\dot{b}\, \beta} =\CP_{\dot{c}\, \gamma}
\epsilon^{\dot{c}\dot{b}} d^{\gamma\beta} \, ,
\quad
\big\{ \, C^{\dot{a}\, \alpha} \, , \,\CP^{\dot{b}\, \beta} \, \big\}
= \epsilon^{\dot{a}\dot{b}} d^{\alpha_\beta}.
$$
Finally, we have the following complete set of canonical variables:
$$
\Gamma^A = (q^i, p_j;\ C^{\dot{a}\, \alpha}, \CP_{\dot{b}\,\beta};\
\lambda^{a\, \alpha}) \, , \qquad a=1,2\, , \quad\dot{a}=1,2\, ,
\quad\alpha = 1,\ldots , m\, .
$$
which includes, besides the original canonical coordinates and
momenta $q^i, \, p_j$, a doublet of canonical ghost--anti--ghost
pairs (\ref{CaPb}), and one canonical pair of the Lagrange multipliers
(\ref{lambda}), for each the constraint pair.

Introduce the generating operators
$\Omega^{\dot{a}\, a} \left( \Gamma\right)$
and ${\CH}\left( \Gamma\right)$ with the following distribution
of the ghost numbers

$$
\varepsilon(\Omega^{\dot{a}\, a})=1\, , \quad
gh(\Omega^{\dot{a}\, a}) = (-1)^{\dot{a}+1} \, , \qquad
\varepsilon({\cal H})=0 \, , \quad gh({\cal H}) = 0 \, .
$$

to be a power series in ghosts, anti--ghosts and Lagrange multipliers
\begin{equation}
\Omega^{\dot{a}\, a}
=C^{ \dot{a} \,\mu} \, T^a_\mu \, - \, \CP^{ \dot{a} \,\alpha }
\lambda^a_\alpha \, + \, \ldots \, , \quad \CH=H+\ldots \, .
\label{boundOH}
\end{equation}
Thus, the pair $\Omega^{\dot{1}\, a} \, , a=1,2$ can
be considered as BRST generators (\ref{boundQH}), while the other
pair $\Omega^{\dot{2}\, a}\, , a=1,2$ should serve as an $Sp(2)$ doublet
of anti-BRST generators.

The Poisson bracket between  $\Omega^{\dot{a}\, a}$  is decomposed
into totally symmetric and antisymmetric part with respect to
permutations of the indices $a=1,2$ and, independently,
$\dot{a}=1,2$:

$$
\big\{\, \Omega^{\dot{a}\, a} \, , \  \Omega^{\dot{b}\, b}\, \big\}
\equiv
\big\{ \,\Omega^{\dot{a}\, a}\, , \ \Omega^{\dot{b}\, b} \, \big\}_{sym} +
\epsilon^{\dot{a}\dot{b}}\epsilon^{{a}{b}}\,\Delta \, ,
$$
where  $\{ \, , \,\}_{sym}$ is a symmetric part:
$$
\big\{\, \Omega^{\dot{a}\, a} \, , \  \Omega^{\dot{b}\, b}
\,\big\}_{sym} \equiv  \frac{1}{4}
\big\{ \, \Omega^{(\dot{a}\,\{ a}\,, \,  \Omega^{\dot{b})\, b\}} \,
\big\}  \, ,
$$
\begin{equation}
\Delta \equiv \frac{1}{4}\,\epsilon_{\dot{a}\dot{b}}\epsilon_{{a}{b}}
\big\{\, \Omega^{\dot{a}\, a}\, ,\,  \Omega^{\dot{b}\, b} \,
\big\} \, , \quad
\varepsilon(\Delta)=0 \, ,  \quad gh(\Delta)= 0 \, .
\label{extdelta}
\end{equation}
We observe, in contrast to the nonextended case, that the Poisson
bracket between the extended BRST generators may involve
antisymmetric part in $a$ and $b$ which emerges when anyone of the
charges is taken together with the anti--charge to another "half" of
the constraints.  As the different constraint "halves" $T^1_\alpha ,
\, T^2_\beta$ do not commute to each other  (\ref{Delta}), the totally
antisymmetric part (\ref{extdelta}) can not vanish.
That is why,  the commutation relations between BRST and anti-BRST
charges have to involve the new generator $\Delta$
(\ref{extdelta})
which has direct analogue neither in a first
 class theory nor in the nonextended BRST formulation
 for a split involution second class system.
 As to the symmetric part of the commutator between the generators
$\Omega^{\dot{a}\, a}$, the formalism contains no Sp(2) covariant
tensors to be involved in the r.h.s. Moreover,
comparison to the nonextended theory reveals obstructions to
noncommutativity in the symmetric sector.

With the account of all said above, one may impose the following
extended system of generating equations
\begin{equation}
\big\{ \, \Omega^{\dot{a}\, a} \, , \,  \Omega^{\dot{b}\, b}  \,
\big\}_{sym} = 0 \, ,
\label{OO}
\end{equation}
\begin{equation}
\big\{ \,
\Omega^{\dot{a}\, a} \, , \,  {\cal H} \, \big\} = 0 \, ,
\label{OH}
\end{equation}
which reveals an explicit $Sp(2)$ symmetry under
ghost--antighost  $Sp(2)$--rotations, and independently, it has
an invariant form with respect to the $Sp(2)$ transformations of
the split involution constraint basis $T^a_\alpha$.
The equations (\ref{OO}), (\ref{OH}) should be supplemented by the
boundary conditions following the definitions (\ref{boundOH})
\begin{equation}
{\left.\frac{\delta\Omega^{\dot{a}\, a}}
{\delta C^{\dot{b}\,\alpha}}\right\vert}_{C={\cal P}=\lambda=0}=
T_\alpha\delta^{\dot{a}}{}_{\dot{b}} \, , \quad
{\left.\frac{\delta\Omega^{\dot{a}\, a}}{\delta \lambda^{b\,\alpha}}
\right\vert}_{C=\lambda=0}= -{\cal P}^{\dot{a}}_\alpha\delta^{a}{}_{b}=
-{\cal P}_{\dot{b}\,\alpha}\epsilon^{\dot{b}\dot{a}}\,\delta^{a}{}_{b}
\, ,
\label{boundO}
\end{equation}

\begin{equation}
{\left.
{\cal H} \right\vert}_{C={\cal P}=\lambda=0}=
 H.
\label{boundH}
\end{equation}

In principle, one may omit the symmetrization in the left hand side
of the equation (\ref{OO}), including simultaneously $\Delta$
(\ref{extdelta}) into the r.h.s. Then the generating equations
(\ref{OO}) read
\begin{equation}
\big\{ \, \Omega^{\dot{a}\, a} \, , \,  \Omega^{\dot{b}\, b}  \,
\big\} = \epsilon^{\dot{a}\dot{b}}\epsilon^{{a}{b}}\,\Delta  \, ,
\label{OOD}
\end{equation}
and involve, in a sense, the revertible element  of the constraint
Poisson bracket matrix $\Delta_{\alpha\beta}$ (\ref{Delta}) entering
$\Delta$ (\ref{extdelta}). This unusual mixture of the revertible
element and involution relations, being encoded in the equations
(\ref{OOD}), contains no contradiction as the compatibility
condition
\begin{equation}
\big\{ \, \Omega^{\dot{a}\, a} \, , \, \Delta \, \big\} = 0
\label{OmD}
\end{equation}
is automatically fulfilled by virtue  of the definition $\Delta$
(\ref{extdelta}) and Jacobi identity. In fact, the totally
antisymmetric part of the first of the relations (\ref{OOD}) is not
a new equation but just the definition for the quantity $\Delta$.
Mention that the equations (\ref{OH}) and (\ref{OOD}) imply the
compatibility condition
\begin{equation}
\big\{ \, \CH \, , \, \Delta
\, \big\} = 0 \, ,
\label{HaD}
\end{equation}
which is not a new restriction as well.
Relations (\ref{OOD},\ref{OmD},\ref{HaD}) show that the split
involution BRST - anti - BRST algebra is extended by
the central element $\Delta$ (\ref{extdelta}), which emerges from the
 noncommutativity of the second class constraints.

The absence of the algebraic contradictions among the generating
equations (\ref{OO}) (or (\ref{OOD})) and (\ref{OH}) does not
automatically provide the existence for the solution with regard of
the given boundary conditions (\ref{boundO}), (\ref{boundH}).
The question is: whether the split involution relations, being imposed
onto the original constraints and Hamiltonian (\ref{split}),
(\ref{splitH}), are sufficient to provide the solution to (\ref{OO}),
(\ref{OH}), or one should strengthen the restrictions to the
constraint algebra.  In general, the problem remains to be
investigated, but for the case of an algebra (more precisely, when
the structure functions $ U^{a}_{\alpha\beta}{}^\gamma$
(\ref{split}), $V_\beta{}^\gamma$ (\ref{splitH}) are  constants) we
have obtained an explicit solution to the generating equations.
As is shown below in this Section, the solution exists for a general
split involution algebra (\ref{split})  and no restrictions appear to
the structure constants $ U^{a}_{\alpha\beta}{}^\gamma$ besides the
respective Jacobi identity (\ref{Jac}). However, the group of the
BRST-anti-BRST symmetric generating equations for the Hamiltonian
$\CH$ (\ref{OH}) restricts the structure constants
$V_\beta{}^\gamma$ (\ref{splitH}) to obey an extra symmetry
requirement. This new condition (which is necessary to provide an
explicit BRST-anti-BRST symmetry) does not restrict the Hamiltonian
on the constraint surface, and could be always met by adding to $H$
certain contributions squared in constraints.  In a sense, this
symmetry requirement to structure constants $V_\beta{}^\gamma$
(\ref{splitH}) is an analog to the restriction imposed to the gauge
fixing condition to provide the BRST-anti-BRST symmetric description
in a first class theory \cite{BLT1,BLT2,Hull,HenGreg}.

Let us expose the solution to generating equations with the split
involution structure coefficients
$ U^{a}_{\alpha\beta}{}^\gamma , \,
 V_\beta{}^\gamma$(\ref{split}), (\ref{splitH}) being constants:
\begin{eqnarray}
\Omega^{\dot{a}\, a}&=& C^{\dot{a}\,\alpha}T^a_\alpha \, \, - \, \,
{\cal P}^{\dot{a}}_\alpha \lambda^{a\,\alpha} \, \, + \,\,
\frac{1}{2}C^{\dot{a}\alpha} C^{\dot{b}\beta} {\cal P}^{\dot{c}}_\nu
\epsilon_{\dot{b}\dot{c}} \, U^a_{\alpha\beta}{}^\nu \, \,  +
 \nonumber\\[4mm]
&&\mbox{}
+ \,\, \frac{1}{2}C^{\dot{a}\,\nu} \lambda^{a\,\alpha}
\lambda^{c}_\beta \epsilon_{cd}  U^{d}_{\nu\alpha}{}^\beta \, \, +
 \,\, \frac{1}{24}C^{\dot{b}\,\alpha} C^{\dot{c}\,\beta}
C^{\dot{a}\,\nu} \lambda^c_\rho
\epsilon_{\dot{b}\dot{c}}\epsilon_{cd}\, U^{(a}_{\alpha\nu}{}^\gamma
U^{d)}_{\gamma\beta}{}^\rho
\, \, +
\nonumber \\[4mm]
&& \mbox{}+
C^{\dot{b}\,\alpha} C^{\dot{c}\,\beta} C^{\dot{a}\,\nu}
\lambda^{a\,\rho} \epsilon_{\dot{c}\dot{b}}\,
Z_{\alpha\beta\nu\rho} \, \, + \, \,
C^{\dot{b}\,\alpha}
C^{\dot{c}\,\beta} C^{\dot{k}\,\mu} C^{\dot{l}\,\nu} C^{\dot{a}\,\rho}
\epsilon_{\dot{b}\dot{c}}\epsilon_{\dot{k}\dot{l}}\,
R^{a}_{\alpha\beta\mu\nu\rho} \,  ,
\label{Oalg}\\[4mm]
{\cal H} \,& =& \, H \, + \, C^{\dot{a}\,\alpha} {\cal
P}^{\dot{b}}_{\beta}V_{\alpha}{}^{\beta}\epsilon_{\dot{a}\dot{b}}
\, + \,
\frac{1}{2}
\lambda^{a\,\alpha} \lambda^{b}_{\beta}\,
V_{\alpha}{}^{\beta}\epsilon_{ba} \, .
\label{Halg}
\end{eqnarray}
The coefficients $ U^{a}_{\alpha\beta}{}^\gamma$ (\ref{Oalg}) are
identified to the respective constants of the split involution
(\ref{split}) and required to obey only the identity (\ref{Jac}).
These constants have the natural symmetry property
\begin{equation}
U^{a}_{\alpha\beta}{}^\gamma = \, -  \,\, U^{a}_{\beta\alpha}{}^\gamma
\, .
\label{Usym}
\end{equation}
The constants
$V_{\alpha}{}^{\beta}$ (\ref{Halg})
should coincide to the
corresponding structure coefficients of the split involution
relations for the original Hamiltonian (\ref{splitH}) and, besides
that, they should be antisymmetric
\begin{equation}
V_{\alpha\beta}\, =\, - \, \, V_{\beta\alpha} \, , \quad
V_{\alpha\beta}=V_{\alpha}{}^{\gamma}\, d_{\gamma\beta} \, .
\label{Vsymmetry}
\end{equation}
In contrast to (\ref{Usym}), the relation (\ref{Vsymmetry}) does not
automatically follow from the respective split involution relations
(\ref{splitH}), so it is a restriction to the definition of the
Hamiltonian $H$ off the constraints.

The higher order structure coefficients  $Z_{\alpha\beta\nu\rho}$ and
$R^{a}_{\alpha\beta\mu\nu\rho}$ (\ref{Oalg}) are expressed via
the original structure constants $U^{a}_{\alpha\beta}{}^\gamma $
as follows
\begin{eqnarray}
Z_{\alpha\beta\nu\rho}=\epsilon_{bc} \Bigg\{ \frac{1}{24}
U^{b}_{\nu\alpha}{}^{\gamma} U^{c}_{\gamma\beta}{}^{\sigma}
d_{\sigma\rho} + \frac{1}{12} U^{b}_{\nu\alpha}{}^{\gamma}
U^{c}_{\gamma\rho}{}^{\sigma}d_{\sigma\beta}+ \nonumber\\[4mm]
\mbox{}+\frac{1}{12} U^{b}_{\beta\rho}{}^{\gamma}
U^{c}_{\gamma\alpha}{}^{\sigma}d_{\sigma\nu}
+ \frac{1}{6} U^{b}_{\alpha\rho}{}^{\gamma}
U^{c}_{\gamma\nu}{}^{\sigma}d_{\sigma\beta} + \frac{1}{8}
\,U^{b}_{\rho\beta}{}^{\gamma} U^{c}_{\nu\alpha}{}^\sigma d_{\sigma\gamma}
\Bigg\};
\label{Z}
\end{eqnarray}
\begin{eqnarray}
R^{a}_{\alpha\beta\mu\nu\rho}=\epsilon_{bc}\bigg[
\ \frac{1}{48}
U^b_{\alpha\mu}{}^{\gamma}
U^c_{\gamma\nu}{}^{\sigma}
U^{a}_{\rho\beta}{}^\theta d_{\theta\sigma}
%%%%%%%%%%%%%%%%%%%%%%%%%%%
+\frac{1}{16} U^{b}_{\rho\nu}{}^{\gamma}
U^a_{\gamma\beta}{}^{\sigma}
U^{c}_{\mu\alpha}{}^\theta d_{\theta\sigma} +
\nonumber\\[4mm]
%%%%%%%%%%%%%%%%%%%%%
+\frac{1}{24}        U^{b}_{\mu\alpha}{}^{\gamma}
                           U^c_{\gamma\rho}{}^{\sigma}
                           U^a_{\sigma\beta}{}^{\theta}d_{\theta\nu}
%%%%%%%%%%%%%%%%%%%%%%%%%%%%%%%%%%
 +\frac{1}{24} U^b_{\rho\mu}{}^{\gamma}
                        U^c_{\gamma\nu}{}^{\sigma}
                        U^a_{\sigma\alpha}{}^\theta d_{\theta\beta}
\bigg].
\label{R}
\end{eqnarray}
In comparison to the respective solution of the split involution
generating equations without an explicit anti-BRST
symmetry (\ref{QQQH})  \cite{split1}, we observe that the terms
 are engaged
of a
new type. In particular, the solution for
$\Omega^{\dot{a} a}$ (\ref{Oalg}) includes contributions of higher
orders in ghosts $C^{\dot{a}\alpha}$ and Lagrange multipliers
$\lambda^{a}_\alpha$.  Of course, the solution (\ref{Oalg}) obeys the
nonextended equations (\ref{QQQH})  with the identification
$\Omega^{\dot{1} a} \equiv Q^a$ , but for $Q^a$ one may find another
solution without higher orders in ghosts and Lagrange multipliers.
Usually only these simplest solutions are considered
in the nonextended case, although the
 general solution to (\ref{QQQH}) contains an arbitrary part
\cite{split3} which may include more complicated terms like those
constitute the higher order contributions in (\ref{Oalg}).

To compare the solution (\ref{Oalg}) with the rank--1 gauge theory in
the BRST--anti--BRST covariant formalism \cite{BLT1}, one may fix
index $a$, choosing in a sense "one half" of the second class
constraints $T^a_\alpha$, \, $a=1,2; \, \alpha=1,\ldots,m$ to be a first
class system and forgetting another "half". In this way,
the respective "half" of the generating equations (\ref{OO}) reduces
to the $Sp(2)$ covariant BRST-anti-BRST equations of a first class
theory \cite{BLT1}. These equations admit more simple solution with
vanishing structure coefficients $Z_{\alpha\beta\nu\rho}$ (\ref{Z})
and $R^{a}_{\alpha\beta\mu\nu\rho}$ (\ref{R}) \cite{BLT1}.  Mention
that solution (\ref{Oalg}) may contain also some unusual terms of
lower orders in ghost and Lagrange multiplier canonical variables.
For example, the first BRST-anti-BRST charge doublet
$\Omega^{\dot{a} \, 1}$ in (\ref{Oalg}),
being related to the first half of the constraints $T^1_\alpha$,
contains contributions like
$$
-\frac{1}{2}C^{\dot{a}\nu}\lambda^{1\alpha}\lambda^{1}_\beta
U^{2}_{\nu\alpha}{}^\beta \, - \,
\frac{1}{24}C^{\dot{b}\alpha}C^{\dot{c}\beta}C^{\dot{a}\nu}
\lambda^{1}_\rho \epsilon_{\dot{b}\dot{c}}
U^{(1}_{\alpha\nu}{}^\gamma
U^{2)}_{\gamma\beta}{}^\rho
$$
which includes structure constant $U^{2}_{\alpha\beta}{}^\gamma$
related to the constraints $T^2_\alpha$.
Due to these terms, the respective charge doublet must have
nontrivial form even for the abelian constraints $T^1_\alpha$.

For the case of one abelian and one non--abelian algebra, e.g. $ U^1=0,
\ U^2\neq 0$ the solution (\ref{Oalg}) for $\Omega^{\dot{a}\,1}$
takes the form
\begin{eqnarray}
\Omega^{\dot{a}\, 1}=
C^{\dot{a}\,\alpha}T^1_\alpha- {\cal
 P}^{\dot{a}}_\alpha\lambda^{1\,\alpha}
%%%%%%%%%%%%%%%%%%%%%%%%%%%%%%%%%
+\frac{1}{2}\,C^{\dot{a}\,\nu} \lambda^{1\,\alpha} \lambda^{1}_\beta\,
   \, U^2_{\alpha\nu}{}^\beta\, .
\label{Orank1/2}
\end{eqnarray}
The second BRST-anti-BRST doublet reads
\begin{eqnarray}
\Omega^{\dot{a}\, 2}= C^{\dot{a}\,\alpha}T^2_\alpha-
 {\cal P}^{\dot{a}}_\alpha\lambda^{2\,\alpha}+
%%%%%%%%%%%%%%%%%%%%%%%%%%%%%%%%%
\frac{1}{2}C^{\dot{a}\,\alpha} C^{\dot{c}\,\beta}
{\cal P}^{\dot{d}}_\nu \epsilon_{\dot{c}\dot{d}}
 \, U^2_{\alpha\beta}{}^\nu +
\nonumber \\[4pt]
\mbox{} + \frac{1}{2} C^{\dot{a}\,\nu} \lambda^{2\,\alpha} \lambda^{1}_\beta\,
U^2_{\alpha\nu}{}^\beta +
\frac{1}{12} C^{\dot{c}\,\alpha} C^{\dot{d}\,\beta} C^{\dot{a}\,\nu}
\lambda^1_\rho \epsilon_{\dot{c}\dot{d}}\,
U^2_{\nu\alpha}{}^\gamma
U^2_{\gamma\beta}{}^\rho\, .
\label{Orank1}
\end{eqnarray}
It is the later charge which coincides to the conventional
rank-1 solution of the first class theory \cite{BLT1}. In this sense,
the second class BRST-anti-BRST theory includes the respective first
class  formulation as a special  split involution case
related to one abelian and one non--abelian algebra.

\section{Unitarizing Hamiltonian}

In previous section, we have formulated generating equations for the
BRST-anti-BRST generators and the Hamiltonian $\CH$. Now we are in
the position to discuss the invariant total Hamiltonian which has to
involve the terms with Lagrange multipliers and constraints, terms
like Faddeev-Popov matrix, etc.

Consider the value $\Delta$ (\ref{extdelta}) which corresponds
the particular solution (\ref{Oalg}) related to the BRST-anti-BRST
generators with the abelian constraints $U^1=U^2=0$:  $$
\Omega^{\dot{a}\, a}= C^{\dot{a}\,\alpha}T^a_\alpha - {\cal
P}^{\dot{a}}_\alpha\lambda^{a\,\alpha}\, , $$ $$ \Delta \equiv
\frac{1}{4}\,\epsilon_{\dot{a}\dot{b}}\epsilon_{{a}{b}} \big\{
\Omega^{\dot{a}\, a},\  \Omega^{\dot{b}\, b} \big\}=
-\frac{1}{4}C^{\dot{a}\,\alpha}
C^{\dot{b}\,\beta}\epsilon_{\dot{a}\dot{b}}\Delta_{\alpha\beta}+
\epsilon_{ab}T^a_\alpha\lambda^{b\beta} -
\frac{1}{2}{\cal P}^{\dot{a}}_\beta{\cal
P}^{\dot{b}}_\alpha \epsilon_{\dot{a}\dot{b}} d^{\alpha\beta}\, .
$$
One may see that the later expression contains all the necessary
"gauge--fixing" terms. As $\Delta$ commutes to
$\Omega^{\dot{a} \, a}$ in general case (\ref{OOD}) (with different
nonvanishing structure functions $U^1, \, U^2$), it could be used to
construct the sought--for unitarizing Hamiltonian in the following
way:
\begin{equation}
H_{complete} = {\cal H}  + \Delta \, ,
\label{Hcomplete}
\end{equation}
the Hamiltonian $\CH$ is defined by equations (\ref{OH}).
In the rank-1 theory (more precisely when all the structure
functions $U^1, \, U^2 , \, V$ are constants), the explicit
expression for $\CH$ has been obtained in the previous section
(\ref{Halg}).
In principle, one can use in the definition (\ref{Hcomplete})
an arbitrary function $F(\Delta)$
instead of $\Delta$ ,
\begin{equation}
H_{complete} = {\cal H}  + F(\Delta) \, .
\label{HF}
\end{equation}
with the only
requirement that $F$ should contain nonvanishing linear part.
Curious to mention that $\Delta$ conserves with respect to this
total Hamiltonian,
$$
\{ \, \Delta \, , \, H_{complete} \, \}
= 0 \, ,
$$
Notice that the definition of the unitarizing Hamiltonian  in
the  BFV--BRST theory involves an arbitrary function
(gauge Fermion $\Psi$ \cite{BFV}, or gauge boson $B$ in the
BRST-anti-BRST symmetric description \cite{BLT1})
which is responsible for the gauge fixing conditions.
In the split involution theory for the second class constraints
\cite{split1,split2},  no any gauge invariance present,
nevertheless, the Hamiltonian $H_{complete}$ (\ref{QQB}) includes
an arbitrary boson $B$ of a ghost number $-2$, and the physical
quantities do not depend on the particular choice of $B$.
In the BRST-anti-BRST symmetric description, the split involution
formalism allows more narrow arbitrariness in unitarizing
Hamiltonian, all the freedom reduces to the choice of a  particular
function $F(\Delta)$ to be introduced into $H_{complete}$ (\ref{HF}).
The wide symmetry of the formalism with 4 odd BRST or anti-BRST
generators makes almost unambiguous the choice of the
invariant unitarizing Hamiltonian.

In the case when one has abelian algebra for one set of the
constraints (say $T^1_\alpha$), and non--abelian for another one, the
charges have the form (\ref{Orank1/2},\ref{Orank1}). The respective part of the
unitarizing Hamiltonian reads
\begin{eqnarray}
\Delta\, = \, -
\frac{1}{4}C^{\dot{c}\,\alpha}C^{\dot{d}\,\beta}
\epsilon_{\dot{c}\dot{d}}
\Delta_{\alpha \beta}+
\epsilon_{cd}T^c_\alpha \lambda^{d\,\alpha} -
\frac{1}{2}{\cal P}^{\dot{c}}_\alpha
{\cal P}^{\dot{d}}_\beta \epsilon_{\dot{c}\dot{d}} d^{\beta\alpha} +
\nonumber\\[4mm]
\mbox{}+
C^{\dot{c}\,\alpha} {\cal P}^{\dot{d}}_\beta
\lambda^{1\,\mu}
\epsilon_{\dot{c}\dot{d}}
 \bigg( \frac{1}{2}U^{2}_{\sigma\alpha}{}^\nu
d^{\sigma\beta}
d_{\nu\mu}
+ U^2_{\mu\alpha}{}^{\beta} \bigg)
+
\frac{1}{2}\lambda^{n\,\alpha}
\lambda^{m\,\beta}\epsilon_{nm}\lambda^1_\nu
U^2_{\alpha\beta}{}^\nu +
\nonumber \\[4mm]
\mbox{} +\frac{1}{8}\, C^{\dot{c}\alpha} C^{\dot{d}\beta}
\lambda^{1}_\mu\lambda^{1}_\nu
\epsilon_{\dot{c}\dot{d}}
\,  U^{2}_{\sigma\alpha}{}^\mu U^{2}_{\beta\gamma}{}^\nu
d^{\gamma\sigma}.
\nonumber
\end{eqnarray}
This expression can be identified to the respective gauge--fixing
contribution in the first class theory if the constraints $T^1_\alpha$
are thought about as gauges. This viewpoint assumes, however, the
extra restriction to the gauges
$ T^1_\alpha = \{ B \, , \, T^2_\alpha \} $
to be abelian.  The latter assumption in fact restricts the
choice of the gauge Boson of the first class theory\footnote{ In
physically relevant cases,  this additional requirement are usually
satisfied by conventional gauges, e.g. in Yang-Mills theory, the
 Lorentz gauge meets the split involution conditions with one abelian
and one non--abelian algebra.  Curious that the split involution
approach allows thereby to introduce two BRST charges and two
respective anti--charges (\ref{Orank1/2}), (\ref{Orank1}) in the
conventional massless Y-M theory. One may consider all these four
independent conserved charges on equal footing.}.

In the general case of 2 different algebras involved in the split
involution relations the "gauge-fixing" part of the
complete Hamiltonian reads
\begin{eqnarray}
&&\Delta\, =
-\frac{1}{4}C^{\dot{c}\,\alpha}C^{\dot{d}\,\beta}\epsilon_{\dot{c}\dot{d}}
\Delta_{\alpha_\beta} +
\epsilon_{cd}T^{c}_\alpha \lambda^{d\,\alpha} -
\frac{1}{2} {\cal P}^{\dot{c}}_\alpha
{\cal P}^{\dot{d}}_\beta \epsilon_{\dot{c}\dot{d}} d^{\alpha\beta} + \nonumber
\\[4mm]
&& \mbox{}+C^{\dot{c}\,\alpha}
{\cal P}^{\dot{d}}_\beta \,
\lambda^{c\,\mu}
\epsilon_{\dot{c}\dot{d}}
\epsilon_{cd} \bigg( \frac{1}{2}U^{d}_{\alpha\sigma}{}^\nu
d^{\sigma\beta}
d_{\nu\mu}
+ U^d_{\alpha\mu}{}^{\beta} \bigg)+
\nonumber\\[4mm]
&&+\frac{1}{2}\lambda^{k\,\alpha}
\lambda^{l\,\beta} \lambda^{c\,\mu}
\epsilon_{kl}
\epsilon_{cd}
U^{d}_{\beta\alpha}{}^\nu d_{\nu\mu}+
\nonumber\\[4mm]
%%%%%%%%%%%%%%%%%%%%%%%%%%%%%%%%%%%%%%%%%%
%%%%%%%%%%%%%%%%%%%%%%%%%%%%%%%%%%%%%%%%%%%
&&+\frac{1}{16}\, C^{\dot{c}\alpha} C^{\dot{d}\beta}
\lambda^{c}_\mu\lambda^{k}_\nu
\epsilon_{\dot{c}\dot{d}}
\epsilon_{cd}\epsilon_{kl}\,
 U^{(d}_{\sigma\alpha}{}^\mu U^{l)}_{\beta\gamma}{}^\nu
d^{\gamma\sigma}
+
\nonumber\\[4mm]
%%%%%%%%%%%%%%%%%%%%%%%%%%%%%%%%%%%%%%%%%%%%%
&&+C^{\dot{c}\,\alpha} C^{\dot{d}\,\beta}
\lambda^{c\,\mu}\lambda^{d\,\nu}
\epsilon_{\dot{c}\dot{d}}
\epsilon_{cd}
\epsilon_{kl}
\bigg(
\frac{1}{8}U^k_{\nu\mu}{}^\gamma   U^l_{\gamma\alpha}{}^\sigma
d_{\sigma\beta}+
\frac{1}{4}U^k_{\alpha\nu}{}^\gamma U^l_{\gamma\mu}{}^\sigma
d_{\sigma\beta}+
\nonumber\\[4mm]
%%%%%%%%%%%%%%%%%%%%%%%%%%%%%%%%%%%%%
&&
\mbox{}+\frac{1}{4}U^k_{\alpha\nu}{}^\sigma U^l_{\beta\mu}{}^\gamma
d_{\gamma\sigma}+ \frac{1}{16}U^k_{\sigma\alpha}{}^\theta
U^l_{\beta\gamma}{}^\lambda  d_{\theta\mu} d^{\gamma\sigma} d_{\lambda\nu}
\bigg)+ \nonumber
\\[4mm]
%%%%%%%%%%%%%%%%%%%%%%%%%%%%%%%%%%%%%%%%%%%%%%%%
&& +C^{\dot{k}\,\alpha} C^{\dot{l}\,\beta}
C^{\dot{c}\,\nu} {\cal P}^{\dot{d}}_\rho
\epsilon_{\dot{k}\dot{l}}\epsilon_{\dot{c}\dot{d}} \bigg(-
Z_{\alpha\beta\nu}{}^\rho + \frac{1}{8} U^k_{\alpha\nu}{}^\gamma
U^{l}{}_{\gamma\beta}{}^\rho\epsilon_{kl} \bigg)+
\nonumber\\[4mm]
%%%%%%%%%%%%%%%%%%%%%%%%%%%%%%%%%%%%%%%%%%%%%%%%%%%%%%%
&& +C^{\dot{k}\,\alpha} C^{\dot{l}\,\beta}
C^{\dot{c}\,\mu} C^{\dot{d}\,\nu}
\lambda^{k\,\rho}
\epsilon_{\dot{k}\dot{l}}\epsilon_{\dot{c}\dot{d}}
\epsilon_{kl}
\bigg(\,\, \frac{1}{2} \big[
R^l_{\rho\mu\alpha\beta\nu}+R^l_{\mu\rho\alpha\beta\nu}+
         R^l_{\alpha\beta\rho\mu\nu}+R^l_{\alpha\beta\mu\rho\nu}\big]-
\nonumber
\\[4mm]
%%%%%%%%%%%%%%%%%%%%%%%%%%%%%%%%%%%%%%%%%%%%%%%%%
         %%%%%%%%%%%%%%%%%%%%%%%%%%%%%%%%%%%%%%%%%%%%%%%%%%
&&
-R^l_{\alpha\beta\mu\nu\rho}
+ \frac{1}{32}\epsilon_{cd} U^c_{\alpha\mu}{}^\gamma
 U^{(d}_{\gamma\nu}{}^\sigma
U^{l)}_{\sigma\beta}{}^\theta d_{\theta\rho}
+ \frac{1}{96}\epsilon_{cd}
U^{(l}_{\alpha\nu}{}^\sigma
 U^{c)}_{\sigma\beta}{}^\gamma
U^d_{\mu \rho}{}{}^{\theta} d_{\theta\gamma} +
\nonumber\\[4mm]
&&+ \frac{1}{4} U^l_{\mu\alpha}{}^\gamma Z_{\gamma\nu\beta\rho}
+ \frac{1}{4} U^l_{\mu\alpha}{}^\gamma
Z_{\nu\gamma\beta\rho}
+\big[ \frac{1}{4}U^l_{\alpha\rho}{}^\gamma +
\frac{1}{2}U^l_{\alpha\sigma}{}^\theta
d^{\sigma\gamma} d_{\theta\rho} \big] Z_{\mu\nu\beta\gamma}
\bigg)+\nonumber\\[4mm]
%%%%%%%%%%%%%%%%%%%%%%%%%%%%%%%%%%%%%%%%
%%%%%%%%%%%%%%%%%%%%%%%%%%%%%%%%%%%
&&+C^{\dot{r}\,\alpha} C^{\dot{s}\,\beta} C^{\dot{c}\,\mu} C^{\dot{d}\,\nu}
C^{\dot{k}\,\gamma} C^{\dot{l}\,\rho} \epsilon_{\dot{r}\dot{s}}
\epsilon_{\dot{c}\dot{d}}\epsilon_{\dot{k}\dot{l}}
\bigg(
\frac{1}{2304}
U^{(c}_{\gamma\alpha}{}^\delta
U^{k)}_{\delta\beta}{}^\lambda
U^{(d}_{\mu\rho}{}^\theta
U^{l)}_{\theta\nu}{}^\sigma
\epsilon_{cd}\epsilon_{kl} d_{\sigma\lambda}-
\frac{1}{2}Z_{\alpha\beta\gamma\sigma} Z_{\mu\nu\rho}{}^\sigma+
\nonumber\\[4mm]
&&\mbox{}+\frac{1}{4} \epsilon_{cd} U^c_{\mu\gamma}{}^\sigma
\big[ R^{d}_{\sigma\nu\alpha\beta\rho}+
R^d_{\nu\sigma\alpha\beta\rho}+
R^d_{\alpha\beta\sigma\nu\rho}+
R^d_{\alpha\beta\nu\sigma\rho} \big]
\bigg).   \nonumber
\end{eqnarray}
This expression is much more cumbersome if compared to the respective
answer of the split involution approach \cite{split1,split2}.
The extra terms of the higher order in ghosts and Lagrange
multipliers must appear to provide the explicit BRST-anti-BRST
symmetry.

Let us discuss in more details the relationship between the
"gauge-fixing" procedure in the split involution theory, being
based on the double commutator (\ref{QQB}), and the respective
expression  $\Delta$  (\ref{extdelta}) of the anti-BRST extended
theory with a single commutation of the charges. With this regard,
the key observation is that the charges $\Omega^{\dot{a} \, a}$, being
solutions of the generating equations (\ref{OO}) with the due ghost
numbers, are linked by the following remarkable relations:
\begin{equation}
\big\{ \Omega^{\dot{a}{a}} \, , \,
B^{\dot{b}\dot{c}} \big\} = \frac{1}{3}\,
\epsilon^{\dot{a} (\dot{b}}\,\,
\Omega^{\dot{c}){a}} \, ,
\label{OBab}
\end{equation}

\begin{equation}
\epsilon_{\dot{c}\dot{b}}\big\{ \Omega^{\dot{b}{a}} \, , \,
B^{\dot{c}\dot{a}} \big\} = \Omega^{\dot{a}{a}} \, ,
\label{OOBab}
\end{equation}
where we have introduced the following Boson function
\begin{equation}
B^{\dot{a}\dot{b}} =\frac{1}{3}{\cal P}^{(\dot{a}}_\alpha
C^{\dot{b})\alpha}\, .
\label{Bab}
\end{equation}
Making use of $B^{\dot{a}\dot{b}}$ (\ref{Bab}) and (\ref{OOBab}), we can
represent $\Delta$ (\ref{extdelta}) as a linear combination of the double
commutators

$$
 \Delta  \equiv
\frac{1}{4}\,\epsilon_{\dot{a}\dot{b}}\epsilon_{{a}{b}} \big\{
\Omega^{\dot{a}\, a},\  \Omega^{\dot{b}\, b} \big\}
  =  - \frac{1}{4}\epsilon_{\dot{a}\dot{b}}
         \epsilon_{\dot{c}\dot{d}}
         \epsilon_{ab}
\big\{ \big\{ B^{\dot{a}\dot{c}}\, , \, \Omega^{\dot{b}a} \big\} ,
 \, \Omega^{\dot{d}b} \big\}.
$$

Taking into account generating equations (\ref{OO}) one finds that
$\Delta$ contains, in fact, only one commutator:
\begin{equation}
\Delta =  \big\{ \Omega^{\dot{1}\, 1},\  \Omega^{\dot{2}\, 2} \big\} =
 -\big\{ \Omega^{\dot{1}\, 2},\  \Omega^{\dot{2}\, 1} \big\}.
\label{nonsymD}
\end{equation}
If one identifies the first
component of the charge-anti-charge doublet
  to be the charge
 $\Omega^{\dot{1}a}\equiv Q^a\,$,
 and considers  $\Omega^{\dot{2}a}$ as the corresponding
anti--charges, the relation (\ref{nonsymD}) will mean that the
"gauge-fixing" part in the unitarizing Hamiltonian (\ref{Hcomplete})
is constituted by the anticommutator between a charge of one half of
the second class constraint set $T^1_\alpha, T^2_\alpha$ and an
anti--charge related to the rest half.
Substituting (\ref{OBab})
into  (\ref{nonsymD}), we obtain different representations for
$\Delta$ in the form of the double commutator
\begin{eqnarray}
\Delta=
\frac{3}{2}\big\{ \big\{ B^{\dot{1}\dot{2}}, \Omega^{\dot{1}a}
\big\}, \Omega^{\dot{2}b}\big\} \epsilon_{ab} = \frac{3}{2} \big\{
\big\{ B^{\dot{1}\dot{2}}, \Omega^{\dot{2}a} \big\},
 \Omega^{\dot{1}b}\big\} \epsilon_{ab}=
 \nonumber \\
= -\frac{3}{4}
\big\{ \big\{ B^{\dot{1}\dot{1}}, \Omega^{\dot{2}a} \big\},
\Omega^{\dot{2}b}\big\} \epsilon_{ab}
 = -\frac{3}{4}
\big\{ \big\{ B^{\dot{2}\dot{2}}, \Omega^{\dot{1}a} \big\},
 \Omega^{\dot{1}b}\big\} \epsilon_{ab}\, .
\nonumber
\end{eqnarray}
Consider the latter noncovariant expression for $\Delta$.
Identifying the ghost-antighost pairs (\ref{CaPb}) to the respective
noncovariant notations,
$$
C^{\dot{1}\alpha}\equiv C^{\alpha}\,,
\,\CP^{\dot{1}\alpha}\equiv - \CP^{\alpha}\,,   \,
\,C^{\dot{2}\alpha}\equiv \BC^{\alpha}\,,
\CP^{\dot{2}\alpha}\equiv \CBP^{\alpha}\ \, ,
$$
we find that $\frac{3}{2}B^{\dot{2}\dot{2}}$ coincides to the gauge-fixing
Boson $B$ in the split involution nonextended formalism (\ref{B}).
$$
\frac{3}{2} B^{\dot{2}\dot{2}} = \frac{1}{2}{\cal
P}^{(\dot{2}}_\alpha C^{\dot{2})\alpha}= \bar{{\cal P}}_\alpha
\bar{C}^\alpha = B  \, ,
$$
and the "gauge-fixing" term $\Delta$ in the unitarizing Hamiltonian
(\ref{extdelta},\ref{Hcomplete}) of the BRST--anti--BRST covariant
theory reduces thereby to the same for as is in the original split
involution theory (\ref{QQB}).
\begin{equation}
\Delta=
\frac{1}{2}\epsilon_{ab} \, \{ \, Q^b \, , \, \{ \,  Q^a \, ,\, B
\, \}\} \, .
\label{deltaQQB}
\end{equation}
Note that the equality (\ref{deltaQQB}) is true only for those
$Q^a \equiv \Omega^{\dot{1}a}\,$ which obey the extended equations
(\ref{OO}), (\ref{OH}), i.e. the unitarizing Hamiltonian of the
nonextended theory (\ref{QQB}) coincides  to the ghost-antighost
covariant expression (\ref{extdelta},\ref{Hcomplete}) only for
a particular solution of the nonextended equations (\ref{QQQH}).
This special solution may differ from the simplest one which has the
lowest possible order in ghosts and Lagrange multipliers.

\section{Examples of the BRST-anti-BRST symmetry \newline
of the second class systems.}

In this section we consider two examples of the nonabelian vector
field models subject to second class constraints. The main question
is to explicitly find the constraint basis obeying the split
involution property. Once this basis has been found, the
BRST-anti-BRST symmetric description is automatically constructed
following the general prescription of the previous two sections of
the paper. In general, the same theory may admit several different
split involution constraint basis. In each of the
examples we choose the basis which is a  simplest one from the
viewpoint of the split involution constraint algebra.  So we choose
two strongly involutive constraint sets for each of the examples,
that provides all extended BRST generators $\Omega^{a\dot{a}}$ to be
linear in the ghost variables.

\subsection{Self--dual nonabelian  model}
Consider $2+1$ nonabelian vector field $f_\mu^A \, , \,\,
\mu=0,1,2;\, A=1,\ldots,N$
with the Lagrangian \cite{TPN}
\begin{equation}\label{1}
L=\frac{1}{2} f_\mu^A f^{\mu\,
A}-\frac{1}{4m}\varepsilon_{\mu\nu\rho} f^{\mu\nu\, A} f^{\rho\, A}
+\frac{1}{12m}\varepsilon_{\mu\nu\rho} c^{ABC} f^{\mu\, A}f^{\nu\, B}
f^{\rho\, C}.
\label{TPNlagr}
\end{equation}
We adopt following notations:
$ \eta_{\mu\nu}=diag(+,-,-)$, $\varepsilon_{012}=1$,
$\varepsilon_{12}=\varepsilon^{12}=1$, $c^{ABC}$ is a total antisymmetric  structure
constant of a Lie group. Covariant derivative and field tensor is
defined as usual:
$$f^{\mu\nu\, A}= \partial^\mu f^{\nu\, A}
-\partial^\nu f^{\mu\, A} + c^{ABC} f^{\mu\, B} f^{\nu\, C}
$$
$$
\nabla^i \phi^A =\partial^i \phi^A + c^{ABC} f^{i\, B}\phi^C.
$$
Introduce momenta $ \pi^A_\mu$ canonically conjugate to
$ f^{\mu \,A}$,
$$
\{f^{\mu\, A} ({\bf x})\, , \, \pi^B_\nu ({\bf y})
\}=\delta^\mu_\nu \delta^{AB} \delta ({\bf x} - {\bf y}).
$$
There are primary constraints
\begin{equation} \label{2} \theta_0^A =\pi_0^A, \end{equation}
\begin{equation} \label{3} \theta_i^A = \pi_i^A
+\frac{1}{2m}\varepsilon_{ij} f^{j\, A}. \end{equation}
The Hamiltonian reads
\begin{equation}
\label{4} H=- \frac{1}{2} f_\mu^A f^{\mu\, A}
+\frac{1}{2m} f^{0\, A} \varepsilon_{ij} f^{ij\,A}.
\end{equation}
Conservation of the primary constraints (\ref{2}),\ref{3}) leads to the
secondary constraints
\begin{equation}
\label{5} \theta^A_3 =f^{0\, A}
- \hat{f}^A,
\end{equation}
where we denote
\begin{equation}
\hat{f}^A =\frac{1}{2m}\varepsilon_{ij} f^{ij\, A}.
\label{fhat}
\end{equation}
The constraints have following Poisson  brackets
$$
\{\theta_0^A(x), \theta_0^B(y)\}=0,
$$
$$
\{\theta_3^A(x), \theta_3^B(y)\}=0,
$$
$$
\{\theta_0^A(x), \theta_3^B(y)\}= -\delta^{AB}\delta^{(2)}(x-y),
$$
$$
\{\theta_i^A(x), \theta_j^B(y)\}=
\frac{1}{m}\varepsilon_{ij}\delta^{AB}\delta^{(2)}(x-y),
$$
$$
\{\theta_i^A(x), \theta_3^B(y)\}= -\frac{1}{m}\varepsilon_{ij}
\left[ \delta^{AB}\partial^j \delta^{(2)}(x-y) +
c^{ABC}f^{j\, C}\delta^{(2)}(x-y)\right].
$$
Thus we observe that all the constraints are of the second class.

It is convenient to rewrite the Hamiltonian in the form
\begin{equation}\label{6}
H_{initial}=\frac{1}{2} f^{i\,A} f^{i\, A}+ \hat{f}^A\hat{f}^A,
\end{equation}
which differs from (\ref{4}) by the contributions
proportional to the constraints $\theta^A_3$.
This Hamiltonian  is in the involution with the constraints
$$
\{\theta_0^A ,H_{initial} \}=0,
$$
$$
\{\theta_3^A, H_{initial} \}=0,
$$
$$
\{\theta_i^A,  H_{initial}\}= -f^{i\, A} -
\frac{1}{m}\varepsilon_{ij}\nabla^j \hat{f}^A.
$$

Constraints $\theta^A_i$ (\ref{3}) are trivial in the sense they
just express the fact that $f^{i\,A}\, , \, i=1,2$ form  canonically
conjugated pairs. Thus we may solve these linear constraints
eliminating momenta $\pi_{i}^{A}$, and define a Poisson
bracket in the respective reduced space\footnote{ This bracket
actually coincides to the Dirac one, if the latter is built of the
constraints $\theta^A_i$ (\ref{3}).}.
The new Poisson brackets read
\begin{equation}
\{f^{i\,A},f^{j\, B} \}= -m\varepsilon^{ij}\delta^{AB} \, .
\label{D}
\end{equation}
Below in this section we use this bracket until mentioned otherwise.
The commutation relations of $\hat{f}^A$ (\ref{fhat}) have the form
$$
\{\hat{f}^A,\hat{f}^B\} =- c^{ABC}\hat{f}^C,
$$
$$
\{\hat{f}^A(x),f^{i\, B}(y)\} =-\delta^{AB}\partial^i\delta^{(2)}(x-y)-
c^{ABC} f^{i\, C}\delta^{(2)}(x-y).
$$
This results in following involution between the constraints
$\theta_0^A\, ,\, \theta_3^A$ (\ref{2},\ref{5}) and the Hamiltonian
\begin{equation}
\begin{array}{rcl}
\{\theta_0^A, \theta_0^B\}&=&0 \, ,\\
[4pt]
\{\theta_0^A, \theta_3^B\}&=& -\delta^{AB} \, ,\\[4pt]
\{\theta_3^A, \theta_3^B\}&=&-c^{ABC}\hat{f}^C  \, ,\\[4pt]
\{\theta_0^A ,H_{initial} \}&=&0  \, ,\\[4pt]
\{\theta_3^A, H_{initial} \}&=&-\partial^i f^{i\, A} \, .
\end{array}
\label{tetainit}
\end{equation}
To meet the split involution conjecture, we choose another basis of
the constraints $T^{aA}$, $a=1,2$, being equivalent to the original one
(\ref{2},\ref{5}), and the equivalent Hamiltonian:
\begin{equation}\label{7}
T^{1A} =\pi_0^A,
\end{equation}
\begin{equation}         \label{8}
T^{2A}= f^{0\, A}-\Phi^{AB} \hat{f}^B,
\end{equation}
\begin{equation}\label{9}
H=\frac{1}{2} (f^{i\, A} -\partial^i \pi_0^B \Phi^{BA})^2+
\frac{1}{2}\hat{f}^A\hat{f}^A,
\end{equation}
where matrix $\Phi^{AB}(\pi_0)$ should satisfy the
Maurer--Cartan equations
(here $\partial^A$ stands for a partial derivative by
$\pi_0^A$)
\begin{equation}   \label{MC} -\partial^A
\Phi^{BC}+\partial^B \Phi^{AC}=c^{A'B'C}\Phi^{AA'}\Phi^{BB'}
\end{equation}
and boundary conditions
\begin{equation}
\Phi^{AB}\vert_{\pi_0=0}=\delta^{AB} \, .
\label{FB}
\end{equation}
These equations has the solution:
\begin{equation}\label{10.1}
\Phi^{AB}=\left[ (e^V-1)V^{-1}\right]^{AB}\, , \qquad V^{AB}=c^{ACB}\pi_0^C.
\end{equation}
This new constraint basis (\ref{7},\ref{8}) and Hamiltonian
(\ref{9}), being apparently equivalent to the original counterparts
(\ref{2},\ref{5}) and (\ref{6}), satisfies the split involution
conjecture (\ref{split}, \ref{splitH}) with vanishing structure
constants:
\begin{equation}
\{T^{1A},T^{1B}\}=0,\ \ \ \ \{T^{2A},T^{2B}\}=0\, ,
\label{a}
\end{equation}
\begin{equation}
\{T^{1A},T^{2B}\}=-\delta^{AB}
\label{b}
\end{equation}
\begin{equation}
\{T^{1A}, H\}=\{T^{2A}, H\}=0 .
\label{c}
\end{equation}
Thus, in this second class theory, we have found the formulation
which allows to construct the  BRST--anti--BRST covariant description
without any artificial gauging procedure.  The respective extended
BRST symmetry generators have the simplest possible form: they are
linear in ghost variables, as both the constraint algebras are
abelian.
\begin{equation}
\Omega^{\dot{a}\, a}
=C^{ \dot{a} \, A} \, T^{a \, A} - \, \CP^{ \dot{a} \,A }
\lambda^{a \,A}  \, .
\label{TPNO}
\end{equation}
The BRST-anti-BRST invariant complete Hamiltonian, being built
according to the general prescription of the Section 4, has
following explicit form:
\begin{eqnarray}
{\cal H}_{complete}&=&
\frac{1}{2}C^{\dot{a}\,A}C^{\dot{b}\,A}\epsilon_{\dot{a}\dot{b}} -
\frac{1}{2} {\cal P}^{\dot{a}\, A}{\cal P}^{\dot{b}\,A}
 \epsilon_{\dot{a}\dot{b}}+
\frac{1}{2} (f^{i\, A} -\partial^i \pi_0^B \Phi^{BA})^2+
\frac{1}{2}\hat{f}^A\hat{f}^A-\nonumber\\[4mm]
&&\mbox{}-\lambda^{2\,A}\pi_0^A +\lambda^{1\,A}(f^{0\,A}-\Phi^{AB}\hat{f}^B).
\end{eqnarray}

The extended BRST symmetry generators and the Hamiltonian depend on
ghosts in a simplest possible way, however they are not polynomials
in the original variables, because of the function $\Phi^{AB}$
(\ref{10.1}). In this respect, this formulation resembles
another BRST description of the model \cite{TPNCo} recently derived
by means of the traditional idea of embedding the second class theory
into an extended phase space to convert the constraints in a first
class. The advantage of our formulation is that i) it does not break
the explicit self--duality and ii) it is globally equivalent to the
original theory, while the first class embedding coincides to the
original second class model only in a certain gauge which may have a
Gribov horizon as the constraints are essentially nonlinear
functions.  So, this converted formulation may actually contain
Gribov's copies which do not correspond the original second class
model.

The problem of nonpolynomial dependence of the momenta could be
overcome if we did not solve primary constraints $\theta^A_i$
(\ref{3}) eliminating momenta $\pi_i^A$ and reducing the Poisson
bracket to the form (\ref{D}). In doing so, we have to find the split
involution basis for all the constraints $\theta^A_0$ (\ref{2}),
$\theta^A_i$ (\ref{3}), $\theta^A_3$ (\ref{5}), and the Hamiltonian
$H$ (\ref{4}). The respective expressions, as we may see, are
polynomials. Introduce the constraints $T^{a \, \mu}, \, a =1,2,\,$
where $\mu$ is a collective index being a pair $\mu=([1,A], [2,A])$:
\begin{eqnarray}
&T^{1\,[1A]}=\theta_3^A =f^{0\,A}-\hat{f}^A,\nonumber\\[4mm]
&T^{1\,[2A]}=
\theta_1^A+\frac{1}{m}\nabla^2\theta_0^A =
\pi_1^A+\frac{1}{2m}f^{2\, A}+\frac{1}{m}\nabla^2\pi_0^A,\nonumber\\[4mm]
&T^{2\,[1A]}=\frac{1}{m}\theta_0^A=\frac{1}{m}\pi_0^A,\label{dop1}\\[4mm]
&T^{2\,[2A]}=
\theta_2^A-\frac{1}{m}\nabla^1\theta_0^A =
\pi_2^A-\frac{1}{2m}f^{1\, A}-\frac{1}{m}\nabla^1\pi_0^A,\nonumber
\end{eqnarray}
$$
\{T^{(a \mu}, T^{b)\nu}\} = 0 ,  \quad
\{T^{[a \mu}, T^{b]\nu}\} = \frac{2}{m}\delta^{\mu\nu}
\varepsilon^{ab}
$$
The Hamiltonian, being modified by weakly vanishing
contributions,
\begin{eqnarray}
H=
\frac{1}{2} (f^{i\, A} +m\varepsilon^{ia}T^{a[2A]})^2+\nonumber\\[3mm]
+\frac{1}{2}(\hat{f}^A-\nabla^a T^{a[2A]}+
\frac{m}{2}c^{ABC}\varepsilon^{ab}T^{a[2B]} T^{b[2C]}
)^2.\label{dop2}
\end{eqnarray} commutes strongly to all the
constraints $T^{a\,\mu}$.  In this formulation we  have twice more
constraints (therefore more ghosts should be introduced), but the
ghosts enter linearly into the BRST generators and quadratically in
the complete Hamiltonian.  Both Hamiltonian (\ref{dop2}) and BRST
charges, being constructed in the general procedure of the section 4
with constraints (\ref{dop1}),  are polynomials in all the variables.

Thus, the method allows to construct explicitly BRST-anti-BRST
invariant description of the self-dual model without any artificial
gauging procedure.

\subsection{Massive Yang--Mills field.}

In the framework of this method, the basic problem to be solved  for
the BRST-anti-BRST covariant quantization of the theory is in
explicit constructing the constraint basis subject to the split
involution requirements.  For the massive nonabelian Yang--Mills
field,  the respective constraints
$T^a_\alpha$, $a=1,2$ have been found in Ref.  \cite{split1}, where
$T^1_\alpha$ are abelian, and $T^2_\alpha$ generate a nonabelian
algebra, related to this field. In this paper, we suggest another
equivalent constraint basis where both $T^1_\alpha$ and $T^2_\alpha$
are abelian, that simplifies the explicit form of the extended BRST
symmetry generators.

We start with the conventional $d=4$ massive Yang--Mills field
Lagrangian
\begin{equation}\label{11}
L=
-\frac{1}{4}G_{\mu\nu}^A G^{\mu\nu\, A} +
\frac{m^2}{2}A_\mu^A A^{\mu\, A},
\end{equation}
where the following conventions are used: metrics $(+,-,-,-)$,
$f^{ABC}$ --- totally anti-symmetric structure constant of a Lie
group, covariant derivative and field tensor are defined as
$$G_{\mu\nu}^A= \partial_\mu A_\nu^{A} -\partial_\nu A_\mu^A +
f^{ABC} A^{\mu\, B} A_\nu^C$$
$$\nabla^\mu \phi^A =\partial^\mu \phi^A + f^{ABC} A^{\mu\, B}\phi^C.
$$
Introduce canonical momenta $ p^A_\mu, \, \mu=0,1,2,3$,
$$
\{A^{\mu\, A}, p^B_\nu\}=\delta^\mu_\nu \delta^{AB}.
$$
The primary constraints are trivial:
\begin{equation} \label{12}
\theta_1^A =p_0^A.
\end{equation}
The Hamiltonian reads
\begin{equation}\label{13}
H=
\frac{1}{2}p_i^A p_i^A+
\frac{1}{4}G_{ij}^A G^{ij\, A} -
A^{0\, A}\nabla^i p_i^A -
\frac{m^2}{2}A^{0\, A} A^{0\, A}+
\frac{m^2}{2}A^{i\, A} A^{i\, A}.
\end{equation}
Conservation of the primary constraints results in the secondary
ones:
\begin{equation}
\label{14} \theta^A_2 =\nabla^i p_i^A +m^2
A^{0\, A}.
\end{equation}
To start with constructing the split involution constraint basis,
we use another Hamiltonian
\begin{equation}
\label{15}
H_{initial}= \frac{1}{2}p_i^A p_i^A+ \frac{1}{4}G_{ij}^A G^{ij\, A} +
\frac{m^2}{2}A^{i\, A} A^{i\, A}+ \frac{1}{2m^2}(\nabla^i
p_i^A)(\nabla^i p_i^A), \end{equation}
which is equivalent to (\ref{13})  as they  differ from each other
by contributions proportional to the constraints $\theta^A_2$.

The constraints have following Poisson brackets
$$
\{\theta_1^A,
\theta_1^B\}=0, $$ $$ \{\theta_1^A, \theta_2^B\}=-m^2\delta^{AB},
$$
$$
\{\theta_2^A, \theta_2^B\}= f^{ABC}\nabla^i p_i^C,
$$
$$
\{\theta_1^A ,H_{initial} \}=0,
$$
$$
\{\theta_2^A, H_{initial} \}=m^2\partial_i A^{i\,A}.
$$
To bring the theory to the split involution form, consider
following constraints and Hamiltonian:
\begin{equation}\label{16}
T^{1A} =p_0^A,
\end{equation}
\begin{equation}\label{17}
T^{2A} =\Phi^{AB}\nabla^i p_i^B +m^2 A^{0\, A},
\end{equation}
\begin{equation}\label{18}
H=
\frac{1}{2}p_i^A p_i^A+
\frac{1}{4}G_{ij}^A G^{ij\, A} +
\frac{m^2}{2}(A^{i\, A} - \frac{1}{m^2}\partial^i p_0^B \Phi^{BA})^2+
\frac{1}{2m^2}(\nabla^i p_i^A)(\nabla^i p_i^A),
\end{equation}
where $\Phi^{AB}=\Phi^{AB}(p_0)$ satisfies the Maurer--Cartan
equations (\ref{MC}) with $c^{ABC}=m^{-2}f^{ABC}$\\
($\partial^A$ stands for a partial derivative by $p_0$)
 and boundary conditions
$\Phi^{AB}\vert_{p_0=0}=\delta^{AB}$ \, .
The solution has the form   (\ref{10.1})
with $V^{AB}=m^{-2} f^{ABC} p_0^C$.
These constraints and Hamiltonian, being equivalent to the initial
ones, obey the split involution requirements:
$$
\{T^{1A},T^{1B}\}=0,\ \ \ \ \{T^{2A},T^{2B}\}=0,
$$
$$
\{T^{1A},T^{2B}\}=-m^2\delta^{AB}
$$
$$
\{T^{1A}, H\}=\{T^{2A}, H\}=0 .
$$
The extended BRST symmetry generators have the same form  as in the
self--dual model (\ref{TPNO}) where the explicit expressions of the
constraints $T^{a\, A}$ are given by Rel. (\ref{16},\ref{17}).  To
conclude, we write down explicitly the complete unitarizing
Hamiltonian
\begin{eqnarray} {\cal H}_{complete}= \frac{1}{2}p_i^A
p_i^A+ \frac{1}{4}G_{ij}^A G^{ij\, A} + \frac{m^2}{2}(A^{i\, A} -
\frac{1}{m^2}\partial^i p_0^B \Phi^{BA})^2+ \frac{1}{2m^2}(\nabla^i
p_i^A)(\nabla^i p_i^A)- \nonumber \\ -\lambda^{2\,A}p_0^A +
\lambda^{1\,A}(\Phi^{AB}\nabla^ip_i^B+m^2 A^{0\, A}) +
\frac{1}{2} m^2
C^{\dot{a}\,A}C^{\dot{b}\,A}\epsilon_{\dot{a}\dot{b}} - \frac{1}{2}
{\cal P}^{\dot{a}\, A}{\cal P}^{\dot{b}\,A}
 \epsilon_{\dot{a}\dot{b}}
\end{eqnarray}
Although ghosts are free in this Hamiltonian, the nonabelian
interaction algebra reveals itself in the theory through the
nonlinear contributions of the function $\Phi^{AB}$, being a solution
to the Maurer-Cartan equations (\ref{MC}).

\section{Concluding remarks}

Let us discuss some remarkable features of the constructed
formulation of the BRST-anti-BRST invariant second class theory.

First, the suggested extension for the BRST-anti-BRST algebra, being
underlaid by the split involution relations,
involves on equal footing 2 BRST generators and 2
anti-BRST ones, and it should contain a new central element $\Delta$
which appears in the commutator of a charge and another anti-charge.
This central element involves the nonvanishing part of the second
class constraint commutator and, thereby it could be never eliminated
from the algebra. What is more, it is the central element $\Delta$
which determines the unitarizing Hamiltonian $H_{complete}$ along
with $\CH$, being a direct BRST-anti-BRST invariant extension of
the original Hamiltonian $H$.  The only ambiguity in the definition
of $H_{complete}$ is in the choice of a particular function
$F(\Delta)$ which incorporates  $\Delta$ in the Hamiltonian.  It is
more narrow arbitrariness in comparison to the case of a
BRST-anti-BRST covariant formulation of first class theory or the
nonextended split involution formalism. In both the later cases, the
unitarizing Hamiltonian involves gauge Boson which is an arbitrary
function (of a certain ghost number) of any original variables,
ghosts and Lagrange multipliers.

Second, we observe that both BRST and anti--BRST symmetry may appear
in a pure second class constrained theory, having no genuine gauge
invariance. It shows that the common recognition of the BRST
invariance as a residual symmetry of the gauge theory is not quite
exact in the sense that the original second class constraint surface,
being conserved in time, may give rise to the corresponding
conservation of the BRST charges, when the ghosts and Lagrange
multipliers are introduced in the formalism.  Thus, the suggested
formulation enables to employ all the tools of the BRST theory to
analyze the second class constrained systems. In particular in a
second class theory, one may attempt to study anomalies, physical
spectrum and consistency of interactions by means of the BRST
cohomological technique.  Mention that the cohomological approach to
construction of interactions, being applied in a gauge theory
\cite{BarHen}, usually brings rather no--go theorems than new
interactions.  For example, for the massless vector field in $d > 3$,
there are no consistent interactions besides the Yang-Mills one.
However, for the second class constrained case, the answers could be
different.  As is known, the ordinary massive Y-M theory corresponds
to the split involution second class constrained system with one
abelian and one nonabelian algebra \cite{split1}, or it could be
formulated even as a pair of abelian but nonpolinomial constraints,
that have explicitly shown in this paper.  The respective extension
of the massive Y-M field with {\it two different nonabelian algebras}
is still unknown, but the suggested formalism does not reveal any
explicit obstacle to introducing such a new type interaction along
the conventional lines of the cohomological approach to consistently
constructing interactions in a constrained theory (for review of the
approach see \cite{hen}).
Mention that the similar intriguing question about the consistent
interactions for higher spin {\it massive} theories
remains yet unsolved, and the split involution scheme seems to be a
natural tool to construct the interaction in this case.

\vspace{1cm}

{\bf Acknowledgments.}

\noindent
Authors are thankful to I.A.Batalin, M.Henneaux, A.Yu.Segal,
A.A.Sharapov, I.V.Tyutin for useful discussions on the
related topics.

This work is partially supported by INTAS--RFBR joint grant 96-829
and RFBR 98-02-16261. S.L.L. appreciates the support from the
International Soros Scientific Education Program, grant d-98-932.

\end{document}